# A class of singular logarithmic potentials in a box with variety of skin thickness and wall interaction

A. D. Alhaidari

*Saudi Center for Theoretical Physics, Jeddah, Saudi Arabia*

We obtain an analytic solution for a three-parameter class of logarithmic potentials at zero energy. The potential terms are products of the inverse square and the inverse log to powers 2, 1 and 0. The configuration space is the one-dimensional box. Using point canonical transformation, we simplify the solution by mapping the problem into the oscillator problem. We also obtain an approximate analytic solution for non-zero energy when there is strong attraction to one side of the box. The wavefunction is written in terms of the confluent hypergeometric function. We also present a numerical scheme to calculate the energy spectrum for a general configuration and to any desired accuracy.

The dynamics of systems confined in space are usually modeled by wave equations with potential functions in a box with non-penetrable walls. The first problem that an undergraduate student of quantum mechanics is asked to solve is that of a particle in a one-dimensional box where the potential function inside the box is zero [1]. There are very few potential functions in the box that render the wave equation analytically solvable. Table 1 accounts for all the classic ones [2]. Nonetheless, using the tridiagonal representation approach, the particle in a box problem was recently solved for the sinusoidal potential function $V(x) = V_0 \cos(k\pi x/a)$, where $k = 1, 2, ..$ and $a$ is the width of the box [3]. On the other hand, physical situations may exist where at some energy the particle in the box would find a stable position at one side of the box or at either side with non-zero probability of tunneling from one side to the other. Therefore, it is of interest to find a potential model in the box with such properties, which is either exactly solvable or even quasi-exactly solvable. The latter means that the problem is analytically solvable for only part of the energy spectrum [4,5]. In this work, we present a class of three-parameter potential functions in a one-dimensional box with these desired qualities. They are inverse square potential functions with singular logarithmic factors. This class of potentials was encountered ten years ago while exploring the group theoretical foundation of the J-matrix method of scattering [6]. The resulting problem associated with this class is analytically solvable for zero energy[†] (e.g., the potential is quasi-exactly solvable). Moreover, solvability requires that the pure inverse square component of the potential (without logarithmic dependence) must vanish. Under other constraints, we find an approximate analytic solution for non-zero energy. The solution of the problem is simplified by using point canonical transformation that maps the problem into the oscillator problem.

---

[†] Note that zero energy in the box may not be the system's lowest energy. That depends on the location of the potential bottom, which is sometimes bottomless. Moreover, zero energy solutions in confined atomic systems are very important when dealing with weakly bound atoms and molecules.

In the atomic units, $\hbar = m = 1$, we consider the following class of three-parameter logarithmic potentials in a one dimensional box of size $a$

$$V(x) = \frac{1/4}{x^2}\left[A + \frac{B}{\ln(x/a)} + \frac{C}{(\ln(x/a))^2}\right], \quad 0 \leq x \leq a, \tag{1}$$

where $A$, $B$, and $C$ are dimensionless parameters. Note that near $x = 0$, the $A$ component of the potential is the most singular (strongest) followed by the $B$ component then the $C$ component. However, near $x = a$, it is the $C$ component that is the strongest followed by the $B$ component. Below, we will show that the time-independent one-dimensional Schrödinger equation associated with this potential,

$$\left[-\frac{1}{2}\frac{d^2}{dx^2} + V(x) - E\right]\phi(x) = 0, \tag{2}$$

is analytically solvable at zero energy and for positive values of the dimensionless parameter $B$. Thus, we take $B = 1$ and plot in Fig. 1 different potential configurations corresponding to different positive values of the parameters $A$ and $C$. We can identify the following three scenarios:

(1) The potential has one minimum and the walls of the box are thick and repulsive.
(2) The potential box has one minimum near $x = 0$ or near $x = a$ where the skin is very thin with glue-like attraction. However, the skin on the other side of the box is thick and repulsive.
(3) The potential box has two minima; one near $x = 0$ and another near $x = a$ with non-zero probability of tunneling from one side to the other. At both sides, the skin is very thin with glue-like attraction.

Changing the sign of $A$ and/or $C$ will change the potential from repulsive to attractive at $x = 0$ and/or $x = a$, respectively. Figure 2 shows such an example where $C < 0$ to be compared with Fig. 1(a). In condensed matter physics, we can model a semiconductor substrate by the potential box where the sharp potential pockets near the walls could be interpreted as being due to atomic interactions of the surface layer atoms with the atoms of the substrate hence giving rise to bound or short lived *local* resonances. Tunneling between the two sides can be increased if one of the potential pockets is much deeper than the other causing resonant tunneling between the trapped states in the flat well and the bound states near the surface of the substrate. Now, we apply the following "point canonical transformation" that preserves the Schrödinger form of Eq. (2)

$$x = ae^{-r^2}, \quad \phi(x) = \sqrt{r}\, e^{-\frac{1}{2}r^2}\psi(r). \tag{3}$$

This transforms the equation into the following Schrödinger-like second order differential equation in the new variable $r \geq 0$

$$\left[\frac{d^2}{dr^2} - \frac{2C + \frac{3}{4}}{r^2} - (1 + 2A)r^2 + (8Ea^2)r^2 e^{-2r^2} + 2B\right]\psi(r) = 0, \tag{4}$$

which is analytically solvable only for $E = 0$, making it a 3D isotropic oscillator-like problem [1]. We take the following ansatz

$$\psi(r) = r^{2\alpha} e^{-\beta r^2} F(r), \quad \beta > 0. \tag{5}$$

Substituting this in Eq. (4) results in a second order differential equation for $F(r)$, which is that of the confluent hypergeometric function ${}_1F_1\left(\alpha - \frac{1}{2}B + \frac{1}{4}; 2\alpha + \frac{1}{2}; r^2\right)$ [7] provided that $A = 0$, $\beta = \frac{1}{2}$ and

$$2\alpha = \frac{1}{2} \pm \sqrt{2C + 1}. \tag{6}$$

–2–

One can easily verify that the $\lim_{x \to a} \phi(x) = 0$ for all $\alpha > -\frac{1}{4}$. Thus, the + and − sign in Eq. (6) corresponds to $C > -\frac{1}{2}$ and $0 > C > -\frac{1}{2}$, respectively. However, at the other end ($x = 0$) and if the confluent hypergeometric series ${}_1F_1$ does not terminate then the $\lim_{r \to \infty} \psi(r) \sim r^{-B-\frac{1}{2}} e^{\frac{1}{2}r^2}$ [8], which makes the $\lim_{x \to 0} \phi(x) \sim \left[\ln(a/x)\right]^{-B/2}$. Thus, to force this limit to vanish we choose $B > 0$. Figure 3 shows the potential function that is compatible with this case where $A = 0$, $B > 0$ and for two values of $C$; one positive and one negative. Figure 4 is a plot of the wavefunction at zero energy for $A = 0$ and for several values of the parameter $B$. If $B$ is large such that $B \gg |C|$, then the zero energy line becomes high above the potential hump at the middle of the box forcing a higher excited state at zero energy. Figure 5 illustrates this situation by giving the wave function for $A = 0$, $B = 20$, and $C = 0.5$. On the other hand, we can also satisfy the boundary condition at $x = 0$ by forcing the series ${}_1F_1$ to terminate giving $\alpha - \frac{1}{2}B + \frac{1}{4} = -n$, where $n = 0, 1, 2, \ldots$ This also dictates that the potential parameter $B$ be positive but forces it to take only discrete values. These values are called "the potential parameter spectrum" [9]. These are the set of values of the potential parameter $B$ that lead to an exact solution at $E = 0$. In fact, for any non-negative integer $n$, the corresponding value of the parameter $B = 2n + 2\alpha + \frac{1}{2}$ in the potential (with $A = 0$ and $C > -\frac{1}{2}$) result in an energy spectrum containing $E = 0$ at the $n^{\text{th}}$ level from the bottom of the spectrum. For example, taking $B = 4 + 2\alpha + \frac{1}{2}$ will produce an energy spectrum in which $E = 0$ is the third line in the spectrum. Table 2 gives the lowest part of the energy spectra corresponding to $n = 0, 1, 2, 3$. The spectrum is obtained numerically by diagonalizing the finite Hamiltonian matrix representation in a suitably chosen basis (see the Appendix for details). Table 3 gives the energy spectrum for non-zero values of the parameter $A$.

To obtain an analytic solution of the problem for $E \neq 0$, we investigate the term $(8Ea^2) r^2 e^{-2r^2}$ in Eq. (4). If the particle is strongly attracted to the right side of the box ($x = a$) due to a situation similar to that in Fig. 2 but with $C < 0$ and $A > 0$ (which flips Fig. 4 horizontally, i.e. $x \to a - x$), then the probability of finding the particle near $x = a$ is very high. Hence, the particle will linger at or near $x = a$ (equivalently, $r = 0$) where we can make the approximation $e^{-2r^2} \approx 1$. Therefore, this term could be approximated as $(8Ea^2) r^2 e^{-2r^2} = (8Ea^2) r^2$ and we can write Eq. (4) *in the neighborhood of $x = a$* as follows

$$\left[ \frac{d^2}{dr^2} - \frac{2C + \frac{3}{4}}{r^2} - \left(1 + 2A - 8Ea^2\right) r^2 + 2B \right] \psi(r) = 0, \quad x \sim a. \qquad (7)$$

This equation is again analytically solvable using the ansatz (5) giving the same solution but requiring a non-zero energy eigenvalue of $E = A/4a^2$.

In conclusion, we gave an interesting potential model in a box with non-penetrable walls. Each wall could be made independently repulsive or attractive. Moreover, the model could have none, one, or two local potential minima. For the last case, the two minima could be configured close to the box walls such that the particle in the box would find a stable position at either side of the box with non-zero probability of tunneling from one side to the other. We found quasi-exact solutions of the potential under certain constrains. If the inverse square component of the potential vanishes then we obtain analytic solution at zero energy. Moreover, we could also obtain an approximate analytic solution for non-zero energy in the neighborhood of the box wall that corresponds to the most singular logarithmic potential term. In the Appendix, we presented an efficient numerical scheme to calculate the energy spectrum for all parameter values and to any desired accuracy. Finally, we note that in the limit as $a \to \infty$, this 1D problem becomes equivalent to a 3D problem with spherical symmetry in which $x$ stands for the radial coordinate and the $A$ term in the potential becomes the orbital term with $A = 2\ell(\ell+1)$. Moreover, a relativistic extension of the box problem is possible by adding a pseudo-scalar coupling of the form $x^{-1}\left[A' + B'/\ln\left(\frac{a}{x}\right)\right]$ to the free 1D Dirac Hamiltonian. In that case an exact solution could be obtained at $E = \pm mc^2$, where $c$ is the speed of light.

**Acknowledgments**: The author is grateful to Profs. H. Bahlouli, A. Jellah, M. S. Abdelmonem and the rest of the theory group at the Saudi Center for Theoretical Physics, Dhahran who helped in making significant improvements on the original manuscript.

## Appendix A: Calculating the energy spectrum

The states of the system described by the potential shown in Eq. (1) consist only of discrete elements since it is totally confined in space. Thus, the energy spectrum of the system is a discrete set, which could be obtained by diagonalizing the corresponding Hamiltonian matrix. The larger the matrix size and the more appropriate the representation basis, the more accurate the spectrum. The following choice of complete basis set is compatible with the ansatz (5) and satisfies the boundary conditions at $x = 0$ and $x = a$ (i.e., $r = 0$ and $r \to \infty$):

$$\chi_n(r) = \sqrt{\frac{\Gamma(n+1)/a}{\Gamma(n+\mu+1)}} \, r^{\mu+\frac{1}{2}} e^{-y/2} L_n^\mu(y), \tag{A1}$$

where $y = r^2$, $L_n^\mu(y)$ is the associated Laguerre polynomial, and $\mu^2 = 2C + 1$ such that $\mu > -1$. Starting from the integral $\int_0^a \phi H \phi \, dx$, we use the transformation (3) and expansion of $\psi(r)$ in the basis (A1) to obtain the matrix elements of the Hamiltonian as follows

$$H_{nm} = \frac{-1}{2a} \int_0^\infty \chi_n(r) \left[\frac{d^2}{dr^2} - \frac{2C+\frac{3}{4}}{r^2} - (1+2A)r^2 + 2B\right] \chi_m(r) \, dr. \tag{A2}$$

Changing variables from $r$ to $y = r^2$ and using the differential equation of the Laguerre polynomials, their recursion relation and orthogonality property, we obtain the following tridiagonal matrix representation of the Hamiltonian

$$4a^2 H_{nm} = \left[(A+1)(2n+\mu+1) - B\right]\delta_{n,m}$$
$$- A\left[\delta_{n,m+1}\sqrt{n(n+\mu)} + \delta_{n,m-1}\sqrt{(n+1)(n+\mu+1)}\right] \tag{A3}$$

On the other hand, the matrix representation of the constant energy term (i.e., the identity multiplying $E$) is

$$\Omega_{nm} = \sqrt{\frac{\Gamma(n+1)\Gamma(m+1)}{\Gamma(n+\mu+1)\Gamma(m+\mu+1)}} \int_0^\infty y^{\mu+1} e^{-3y} L_n^\mu(y) L_m^\mu(y) \, dy. \tag{A4}$$

After evaluating this integral, we obtain the energy spectrum by solving numerically the generalized matrix eigenvalue equation $H|\psi\rangle = E\Omega|\psi\rangle$ for large enough matrix size. The integral (A4) could be evaluated using the general formula (14) in [10] giving

$$\Omega_{nm} = 3^{-\mu-2}\sqrt{n!m!\Gamma(n+\mu+1)\Gamma(m+\mu+1)}$$

$$\times \sum_{k=0}^{\min(n,m)} (k+\mu+1)\left[3^{2k}\Gamma(k+\mu+1)k!(n-k)!(m-k)!\right]^{-1} {}_2F_1\left(\begin{matrix}-n+k,k+\mu+2\\k+\mu+1\end{matrix}\bigg|\frac{1}{3}\right) {}_2F_1\left(\begin{matrix}-m+k,k+\mu+2\\k+\mu+1\end{matrix}\bigg|\frac{1}{3}\right)$$

where ${}_2F_1\left(\begin{matrix}a,b\\c\end{matrix}\bigg|z\right)$ is the hypergeometric function. Care must be taken when programming this expression in the calculation for large indices *n* and *m* since one may encounter multiplication of very large numbers with very small resulting in reduced accuracy.

**Tables Caption:**

**Table 1**: A list of all classic potential functions in a one-dimensional box that result in an exact solution. *A* and *B* are real dimensionless parameters. The sign of the singularity of the potential at the left and right walls of the box is given by the sign of the parameters listed in the right column.

**Table 2**: The energy spectrum (in units of $1/a^2$) for the potential parameters $A = 0$, $C = 0.1$, $B = 2n + 2\alpha + \frac{1}{2}$, and with 70×70 Hamiltonian matrix. Energy levels designated by stars are too high or too low to obtain at this accuracy.

**Table 2**: The energy spectrum (in units of $1/a^2$) calculated as shown in the Appendix for $B = 1$, $C = 0.2$, and for several values of the parameter *A*. The energy level designated by stars is too high to obtain at this accuracy.

**Table 1**

| Name | $a^2 V(x)$ | Box Space | Singularity Sign [Left, Right] |
|---|---|---|---|
| Pöschl-Teller | $\dfrac{A}{\cos^2(\pi x/a)} + \dfrac{B}{\sin^2(\pi x/a)}$ | $0 \le x \le a/2$ | $[B, A]$ |
| Rosen-Morse | $\dfrac{A}{\cos^2(\pi x/a)} + B \tan(\pi x/a)$ | $-a/2 \le x \le a/2$ | $[A, A]$ |
| Scarf | $\dfrac{A + B\cos(\pi x/a)}{\sin^2(\pi x/a)}$ | $0 \le x \le a$ | $[A+B, A-B]$ |

**Table 2**

| Level | $n = 0$ | $n = 1$ | $n = 2$ | $n = 3$ |
|---|---|---|---|---|
| 0 | 0.000000 | −80.917441 | −********* | −******* |
| 1 | 14.143045 | 0.000000 | −19.535388 | −4577.59 |
| 2 | 144.10743 | 42.370935 | 0.000000 | −14.750982 |
| 3 | 1211.470 | 658.389 | 77.470172 | 0.000000 |
| 4 | 9542 | 6617.0 | 943.581 | 130.71607 |
| 5 | 72700 | 56400 | 8566 | 2158.78 |
| 6 | 540000 | 440000 | 70600 | 30670 |
| 7 | ****** | ****** | 550000 | 306000 |

**Table 3**

| Level | $A = 0.0$ | $A = 0.5$ | $A = 1.0$ | $A = 1.5$ |
|---|---|---|---|---|
| 0 | 3.423315 | 4.696735 | 5.629147 | 6.42104 |
| 1 | 21.914694 | 27.109924 | 31.191909 | 34.77632 |
| 2 | 182.62288 | 213.402057 | 237.673658 | 258.93347 |
| 3 | 1427.901 | 1627.888 | 1787.278 | 1927.078 |
| 4 | 10863 | 12232 | 13338 | 14312 |
| 5 | 80800 | 90200 | 98000 | 104800 |
| 6 | ***** | ***** | 600000 | 670000 |

**Figures Caption:**

**Fig. 1**: Plots of the potential (in units of $1/a^2$) as a function of $x$ (in units of $a$) for $B=1$ and for various values of the parameters $A$ and $C$: (a) $A=0.25$ and $C=0.01$, (b) $A=0.25$ and $C=0.1$, (c) $A=0.5$ and $C=0.01$, (d) $A=0.5$ and $C=0.2$

**Fig. 2**: Plot of the potential (in units of $1/a^2$) as a function of $x$ (in units of $a$) for parameter values $B=1$, $A=0.25$ and $C=-0.01$

**Fig. 3**: Plots of the potential (in units of $1/a^2$) as a function of $x$ (in units of $a$) for $A=0$, $B=1.0$ and for $C=0.01$ (Fig. 3a), $C=-0.01$ (Fig. 3b).

**Fig. 4**: Plot of the wavefunction $\phi(x)$ at zero energy for $A=0$, $C=0.1$, and for $B = 0.5$, 1.0, 1.5, and 2.0

**Fig. 5**: Plot of the wavefunction $\phi(x)$ at zero energy for $A=0$, $B=20$, and $C=0.5$. Zooming in near $x=0$ exposes many oscillations. We found seven nodes of the wave function for $x \leq 0.1a$

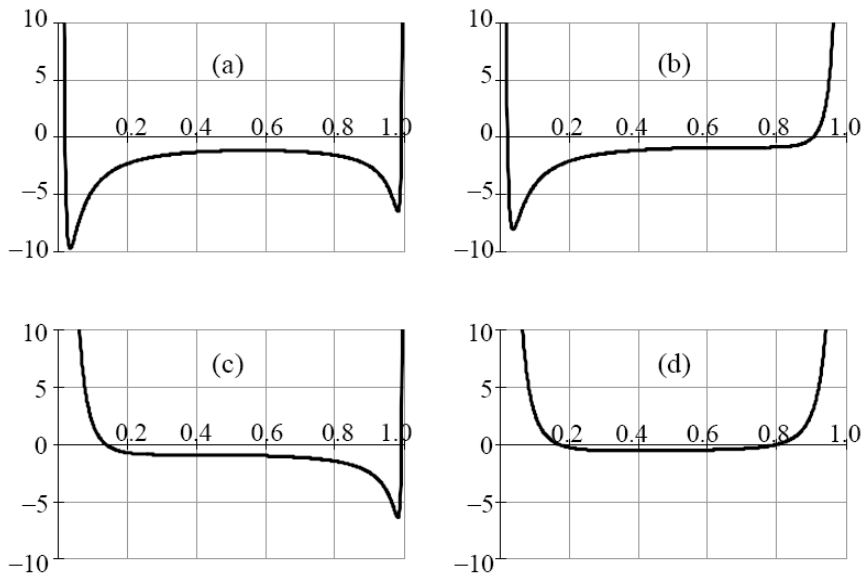

**Fig. 1**

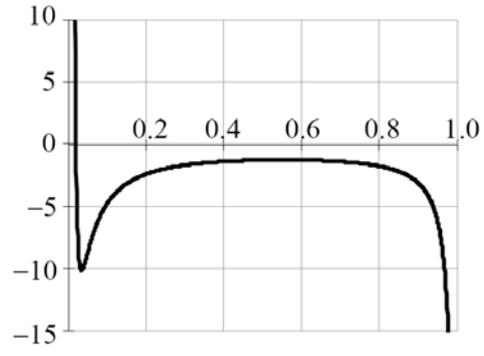

**Fig. 2**

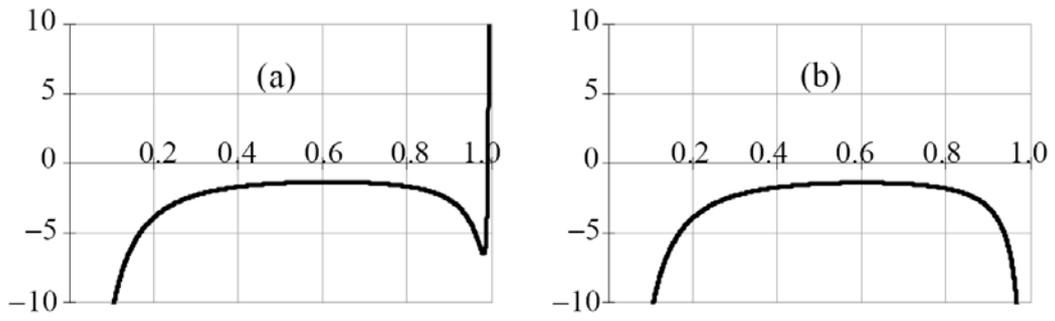

**Fig. 3**

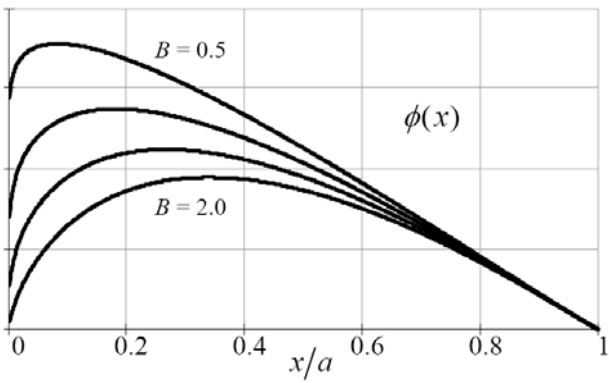

**Fig. 4**

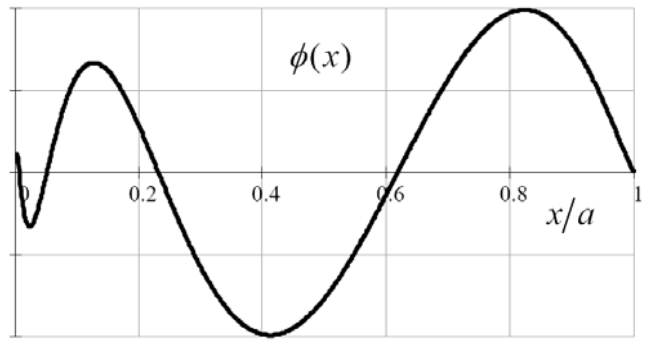

**Fig. 5**